\documentclass{article}

\usepackage{fullpage}
\usepackage{mathptmx}       % selects Times Roman as basic font
\usepackage{helvet}         % selects Helvetica as sans-serif font
\usepackage{courier}        % selects Courier as typewriter font
\usepackage{type1cm}        % activate if the above 3 fonts are
\usepackage{makeidx}         % allows index generation
\usepackage{graphicx}        % standard LaTeX graphics tool
\usepackage{multicol}        % used for the two-column index
\usepackage[bottom]{footmisc}% places footnotes at page bottom

\usepackage{amssymb, amsmath, amsthm, bm}
\usepackage[mathscr]{eucal}

\newcommand{\Z}{{\mathbb Z}}
\newcommand{\R}{{\mathbb R}}
\newcommand{\N}{{\mathbb N}}
\newcommand{\sX}{{\mathscr X}}
\newcommand{\sA}{{\mathscr A}}
\newcommand{\sB}{{\mathscr B}}
\newcommand{\sd}{\Sigma\Delta}
\newcommand{\sigmin}{\sigma_{\mathrm{min}}}
\newtheorem{theorem}{Theorem}[section]
\newtheorem{proposition}{Proposition}[section]
\newtheorem{definition}{Definition}[section]

\makeindex             % used for the subject index
\begin{document}

\title{Noise-shaping Quantization Methods for Frame-based and Compressive Sampling Systems}
\author{Evan Chou\thanks{Courant Institute, 251 Mercer Street, New York, NY 10012 USA and Google New York},\ \ C.~Sinan G\"unt\"urk\thanks{Courant Institute, 251 Mercer Street, New York, NY 10012 USA},\ \  Felix Krahmer\thanks{University of G{\"o}ttingen, Institute for Numerical and Applied Mathematics, Lotzestra\ss e 16-18, 37083 G{\"o}ttingen, Germany},\ \   Rayan Saab\thanks{University of California, San Diego, 9500 Gilman Drive,
Dept. 0112, La Jolla, CA 92093 USA},\ \  \"Ozg\"ur Y{\i}lmaz\thanks{University of British Columbia, 1984
 Mathematics Road, Vancouver, BC V6T 1Z2 Canada} }
\maketitle

\abstract{Noise shaping refers to an analog-to-digital conversion methodology in which quantization error is arranged to lie mostly outside the signal spectrum by means of oversampling and feedback. Recently it has been successfully applied to more general redundant linear sampling and reconstruction systems associated with frames as well as non-linear systems associated with compressive sampling. This chapter reviews some of the recent progress in this subject.}

\section{Introduction}
\label{intro}

Source coding via quantized linear representations, also known as transform coding, is a classical and well-studied subject. Yet it is poorly understood outside the simple setting of orthogonal transforms, namely, for frame-based representations. The same can also be said for partially nonlinear representations such as those based on compressive sampling. The basic reason for the difficulty in solving the quantization problem for these more general sampling and reconstruction systems is the lack of an analog of Parseval's identity which, more or less, dictates the best quantization strategy for orthogonal systems. While some kind of basic reconstruction stability can be ensured relatively easily, these results do not offer correct rate-distortion trade-offs because of their inefficiency in utilizing redundancy, especially under constraints that do not allow for high-resolution quantization.

Redundancy is a key concept of frame-based as well as compressive sampling systems. It can be understood in terms of the sampling process (e.g., what part of the coefficient space is taken up with the actual measurements) or in terms of the reconstruction process (e.g., which perturbations of the measurements have the smallest effect on the reconstruction). Efficient encoding via the first approach is generally not practical because codewords cannot be easily placed arbitrarily in the coefficient space. Indeed, quantized measurements are typically required to lie on a finite rectangular grid. An alternative approach is then to seek ways of arranging the quantization error in the coefficient space to lie in directions that are away from the actual measurements, typically by means of some feedback process. {\em Noise shaping} is the generic name of this quantization methodology. It has its roots in sigma-delta modulation, which is used for oversampled analog-to-digital (A/D) conversion \cite{IY, NST96, Candy-Temes, SchTe04}.

Let us explain the philosophy of noise shaping in more concrete terms.
In both frame-based and compressive sampling systems, we have a linear sampling operator $\Phi$ that can be inverted on a given space $\sX$ of signals using some (possibly nonlinear) reconstruction operator $\Psi$. Given a signal $x\in \sX$ and its sampled version $y=\Phi x$, ordinarily we recover $x$ exactly (or approximately, as in compressive sampling) as $\Psi(y)$. In the context of this paper, quantization of $y$ will mean replacing it with a vector $q$ which is of the same dimensionality as $y$ and whose entries are chosen from some given alphabet $\sA$. The goal is to choose $q$ so that the approximate reconstruction $x^\#:=\Psi(q)$ is as close to $x$ as possible as $x$ varies over $\sX$.

In the context of finite frames, $\Phi$ is a full-rank $m\times k$ matrix where $m > k$, and $\Psi$ is any left inverse of $\Phi$. The rows of $\Phi$ form the {\em analysis frame} and the columns of $\Psi$ form a {\em synthesis frame} dual to this frame. With $y = \Phi x$ and $x = \Psi y$ as above, when $y$ is replaced by a quantized vector $q$, the reconstruction error $e:=x - x^\#$ is equal to $\Psi(y-q)$. Therefore the correct strategy to reduce the size of $e$ is not to minimize the Euclidean norm $\|y-q\|$ as memoryless scalar quantization (MSQ) does, but to minimize the semi-norm $|y-q|_\Psi := \|\Psi(y-q)\|$. 
In other words, we seek $q\in \sA^m$ so that the quantization ``noise'' $y-q$ is close to $\mathrm{ker}(\Psi)$ in the above sense. This is the basic principle of noise shaping. How this goal can be achieved (approximately), i.e., the actual process of noise shaping, as well as what noise shaping can offer for source coding are nontrivial questions that will be addressed in this article.

While the basic principle of noise shaping is formulated above for linear sampling and reconstruction systems, its philosophy extends to compressive sampling systems where the reconstruction operator is generally nonlinear. The simplest connection is made by considering strictly sparse signals. Let $\Sigma^N_k$ denote the nonlinear space of $N$-dimensional vectors which have no more than $k$ nonzero entries. In the context of compressive sampling, $\Phi$ is an $m\times N$ matrix where $m \ll N$, which means that the sampling process is lossy for the whole of $\R^N$. However, note that  $\Sigma^N_k$ is the union of (a large number of) $k$-dimensional linear subspaces on each of which $\Phi$ acts like a frame once $m > k$. This observation opens up the possibility of noise shaping. Indeed, fixing any one of these subspaces $V$, we can envision a noise shaping process associated with any of the linear inverses (duals) of $\Phi$ on $V$. However, it is not clear how one might organize all of these individual 
noise shaping processes, especially given that these subspaces are not directly available to the quantizer. What comes to the rescue is the notion of an {\em alternative dual}. While we formulated noise shaping above as matching the quantization operator to a given dual frame, it is also possible to consider matching the dual frame to a given quantization operator. This results in the possibility of ``universal'' quantization processes (i.e., independent of the signal subspace) which become noise-shaping processes for suitable alternative duals. Even though finding these suitable alternative duals may require extracting information about the signal subspace, this duty purely belongs to the decoder and not the quantizer. 

This article is organized as follows. In Section \ref{classical}, we review the basics of classical noise shaping in the setting of sigma-delta ($\sd$) modulation. In Section \ref{alternative}, we extend the formulation of noise shaping and introduce various notions of alternative duals for noise shaping in the setting of frames, followed by their performance analysis for random frames in Section \ref{random-frames}. We then discuss noise-shaping quantization methods for compressive sampling in Section \ref{CS}.

\section{Classical noise shaping: Sigma-Delta Modulation}
\label{classical}

The Shannon-Nyquist sampling theorem for bandlimited functions provides the natural framework of conventional A/D conversion systems. With the Fourier transform normalized according to the ``ordinary-frequency'' convention 
$$ \widehat x(\xi) := \int_{-\infty}^\infty x(t) e^{-2\pi i \xi t} \,\mathrm{d}t, $$
let us define the space $\sB_\Omega$ of bandlimited functions to be all $x$ in $L^2(\R)$ such that $\widehat x$ is supported in $[-\Omega,\Omega]$. The classical sampling theorem says that any $x \in \sB_\Omega$
can be reconstructed perfectly from its time samples $(x(n\tau))_{n \in \Z}$ 
according to the formula
\begin{equation} \label{shannon}
x(t) = \tau \sum_{n \in \Z} x(n\tau) \psi(t - n\tau),
\end{equation}
where $\tau \leq \tau_\mathrm{crit} := \frac{1}{2\Omega}$, and
$\psi$ is any function in $L^2(\R) $such that 
\begin{equation}\label{admissible}
\widehat \psi(\xi) = \left \{
\begin{array}{ll}
1, & |\xi| \leq \Omega, \\
0, & |\xi| > \frac{1}{2\tau}.
\end{array}
\right.
\end{equation}
Hence, if we define the sampling operator $(\Phi x)_n:= x(n\tau)$ and the reconstruction operator $\Psi(u) := \tau \sum u_n \psi(\cdot - n \tau)$ (on any space it makes sense), then $\Psi$ is a left inverse of $\Phi$ on $\sB_\Omega$  when $\tau$ and $\psi$ satisfy the conditions stated above.

The value $\rho:=1/\tau$ is called the sampling rate, and
$\rho_\mathrm{crit} := 1/\tau_\mathrm{crit} = 2\Omega$ is called the critical
(or Nyquist) sampling rate. Their ratio given by 
\begin{equation}\label{lambda}
\lambda:= \frac{\rho}{\rho_\mathrm{crit}}
\end{equation}
is called the {\em oversampling ratio}. According to the value of $\lambda$,
A/D converters are broadly classified as Nyquist-rate converters ($\lambda \approx 1$) or oversampling converters ($\lambda \gg 1$).

Nyquist-rate converters set their sampling rate $\rho$ slightly above the critical frequency $2\Omega$ so that $\psi$ may be chosen to decay rapidly enough to ensure absolute summability of \eqref{shannon}. Given any quantization alphabet $\sA$, the (nearly) optimal quantization strategy in this (nearly) orthogonal setting is memoryless scalar quantization (MSQ). This means that each sample $y_n := x(n\tau)$ is rounded to the nearest quantization level $q_n \in \sA$. This process is also referred to as pulse-code modulation (PCM). If each sample is quantized with error
no more than $\delta$, i.e., $\|y-q\|_\infty \leq \delta$, then 
the error signal 
\begin{equation} \label{err_sig}
e(t) := x(t) - (\Psi q)(t) = \tau \sum_{n \in \Z} \big (y_n - q_n \big) \psi(t - n\tau)
\end{equation}
obeys the bound 
$\|e\|_{L^\infty} \leq C \delta$ where $C$ is independent of $\delta$.
This is essentially the best error bound one can expect for
Nyquist-rate converters. Because setting $\delta$ very small is
costly, Nyquist-rate converters are not very suitable for signals that
require high-fidelity such as audio signals.

Oversampling converters are designed to take advantage of the redundancy in the representation \eqref{shannon} when $\tau < \tau_\mathrm{crit}$. In this case, the interpolation operator $\Psi$ has a kernel which gets bigger as $\tau \to 0$. Indeed, let $\widehat \psi(\xi) =0$ for $|\xi| > \Omega_0$. 
It is easily seen that $\Psi u = 0$ if 
\begin{equation}\label{KerT}
\sum_{n\in\Z} u_n e^{2\pi i n \xi} = 0 \mbox{ for } |\xi| < \tau\Omega_0.
\end{equation}
This means that even though $y-q$ may be large everywhere, $e=\Psi(y-q)$ can be very small if $y-q$ can be arranged to be  spectrally disjoint from the (discretized) reconstruction kernel $\psi$. This is the concrete form of noise shaping that we briefly discussed in the Introduction.

The main focus of an oversampling A/D converter is on its quantization algorithm, which has to be
non-local to be useful, but also causal so that it can be implemented in real time. The assignment of each $q_n$ will therefore depend on $y_n$ as well as a set of values (the states) that can be kept in an analog circuit memory, while meeting the spectral constraints on $y-q$ as described in
the previous section. $\sd$ modulators operate according to these principles.

\begin{figure}[tp]
\centering
\centerline{\includegraphics[scale=0.44]{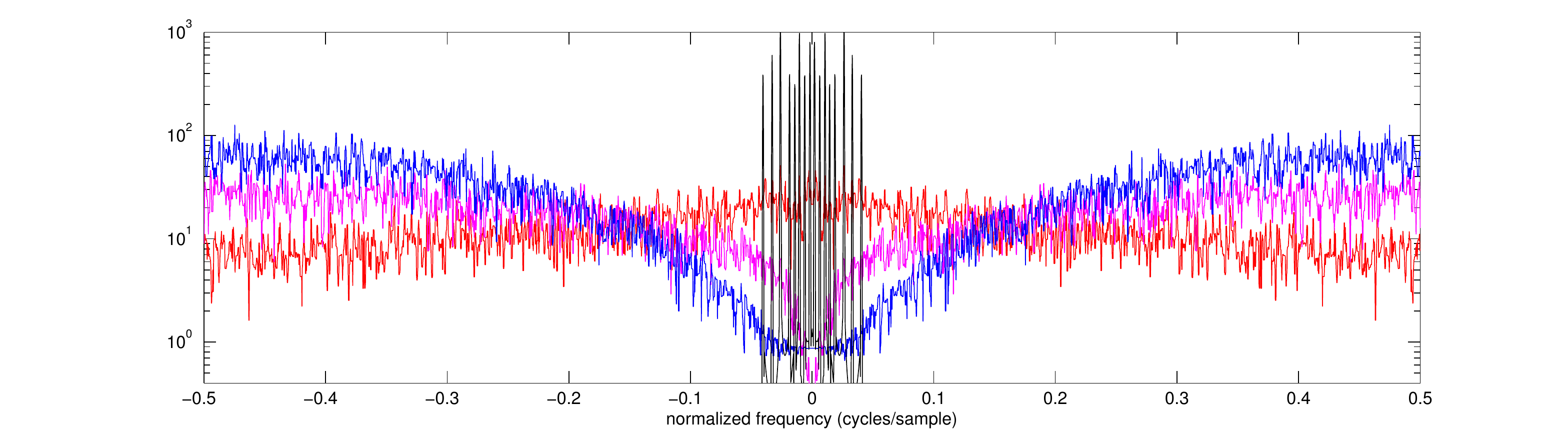}}
\caption{Illustration of classical noise shaping via $\sd$ modulation: The superimposed Fourier spectra of a bandlimited signal (in black), and the quantization error signals using MSQ (in red), 1st order $\sd$ modulation (in magenta), and 2nd order $\sd$ modulation (in blue).}
\label{fig:noise_shaping_illustration}
\end{figure}

As can be seen in \eqref{KerT}, the kernel of $\Psi$ consists of high-pass sequences. Hence the primary objective of $\sd$ modulation is to arrange the 
quantization error $y - q$ to be an approximate high-pass sequence (see Fig. \ref{fig:noise_shaping_illustration}).
This objective can be realized by setting up a difference equation, the 
so-called {\em canonical} $\sd$ equation, of the form
\begin{equation}\label{sig-del}
y-q = \Delta^r u,
\end{equation}
where $\Delta$ denotes the finite difference operator defined by
\begin{equation}\label{def_Delta}
(\Delta w)_n:=w_n - w_{n-1},
\end{equation}
$r$ denotes the ``order'' of the scheme, and $u$ is an
appropriate auxiliary sequence called the {\em state sequence}.
This equation does not imply anything about $q$ without any constraint on
$u$. The most useful constraint turns out to be boundedness.

In practice, the boundedness of $u$ in 
(\ref{sig-del}) has to be attained through a recursive algorithm.
This means that given any input sequence $(y_n)$, the $q_n$ are found by 
a given ``quantization rule'' of the form
\begin{equation}\label{Q_rule}
q_n = F(u_{n-1},u_{n-2},\dots,y_n,y_{n-1},\dots),
\end{equation}
and the $u_n$ are updated via
\begin{equation}\label{u_update}
u_n = \sum_{k=1}^r (-1)^{k-1} \binom{r}{k} u_{n-k} + y_n - q_n,
\end{equation}
which is a restatement of (\ref{sig-del}). In electrical engineering, such a recursive procedure for quantization is called ``feedback quantization" due to the role $q_n$ plays as a feedback control term in (\ref{u_update}). The role of the quantization rule $F$ is to keep the system {\em stable}, i.e., $u$ bounded.

Stability is a crucial property. Indeed, it was shown in \cite{DD} that a stable $r$th order scheme results in the error bound
\begin{equation} \label{err1}
\|e \|_{L^\infty}  \leq  \|u\|_{\ell^\infty} \|\psi^{(r)} \|_{L^1}\tau^r,
\end{equation}
where $\psi^{(r)}$ denotes the $r$th order derivative of 
$\psi$. The implicit $\Omega$- and the explicit $\tau$-dependence of this estimate can be
replaced with a single $\lambda$-dependence by setting
$\psi(t) := \Omega \psi_0(\Omega t)$ where the prototype $\widehat \psi_0(\xi)$
equals $1$ on $[-1,1]$ and vanishes for $|\xi | \geq 1+\epsilon_0$, with
$\epsilon_0>0$ fixed. Let $C_0:=\|\psi_0 \|_{L^1}$. 
Bernstein's inequality applied to $\psi$ yields
\begin{equation} \label{err2}
\|e \|_{L^\infty}  \leq  C_0 \|u\|_{\ell^\infty} \pi^r(1+\epsilon_0)^r
\lambda^{-r}, \mbox{ for all } \lambda > 1+\epsilon_0.
\end{equation}

With this error bound, there are two goals in progression.
The first is to keep $u$ bounded and the second is to keep the bound small. Ultimately,
the best strategy is to have, for each $r$, a quantization rule yielding a 
stable $r$th order scheme, and then for any given $\lambda$, 
to choose the best one (i.e., the one with the least
error bound). This task
is significantly complicated by the fact that the bound on $u$ has a strong dependence on $r$, especially for small quantization alphabets $\sA$. 
In general it is not possible to expect this dependence to be less than $(cr)^r$ for some constant $c$ that depends on the given amplitude range $\mu$ for $x$.
This growth order is also what is needed to ensure that the reconstruction error decays exponentially, i.e., as $2^{-p\lambda}$, as a function of $\lambda$, which is the best possible due to Kolmogorov entropy estimates for bandlimited functions \cite{exp_decay}. The rate $p$ of exponential decay that is achievable by the resulting family of schemes is inversely proportional to $c$, and gets worse as $\mu$ is increased. The question of best achievable accuracy for oversampling converters in this setting remains open. Currently, the best result in the one-bit case with $\sA = \{-1,1\}$ yields $\|e\|_{L^\infty} = O(2^{-p\lambda})$ where $p = \pi/(6e^2 \log 2) \approx 0.1$, and $\mu \approx 0.06$. Higher values of $p$ can be achieved with more levels in $\sA$. For example, if $\sA=\{-1,0,1\}$, then $p$ rises to $0.15$ and $\mu$ to $0.25$ \cite{DGK}. These are rigorously proven bounds and the actual behavior of the error based on numerical experiments appears to be better. For the details of the quantization 
rules which result in these exponentially accurate $\sd$ modulators, see \cite{exp_decay, DGK}. It has also been shown that no matter how the bits are assigned the rate of the exponential decay cannot match that of Nyquist-rate conversion \cite{KW}. 

\section{Generalized Noise-shaping Operators and Alternative Duals of Frames for Noise Shaping}
\label{alternative}

In this section, we will generalize the classical theory of $\sd$ modulation to more general noise-shaping quantizers as well as sampling and reconstruction systems. For conceptual clarity, we will separate the process of noise shaping from the processes of sampling and reconstruction. While we will present these generalizations in a finite-dimensional setting, extensions to infinite-dimensional settings are usually possible. We will also discuss the notion of alternative duals of frames which are associated with noise-shaping quantizers. 

\subsection{A general framework of noise shaping}
The canonical $\sd$ equation we saw in \eqref{sig-del} is a special case of a more general framework of noise shaping. Let $\sA$ be a finite quantization alphabet and $J$ be a compact interval in $\R$. Let $h = (h_j)_{j\geq0}$ be a given sequence, finite or infinite, where $h_0=1$. By a noise-shaping quantizer with the transfer filter $h$, we mean any sequence $Q=(Q_m)_1^\infty$ of maps $Q_m:J^m \to \sA^m$, $m \in \N$, where for each $y \in J^m$, the output $q := Q_m(y)$ satisfies
\begin{equation} \label{y-q-h-u}
y - q = h*u 
\end{equation}
where $u \in \R^m$ and $\|u\|_\infty \leq C$ for some constant $C$ which is independent of $m$.
Here $h*u$ refers to the (finite) convolution of $h$ and $u$ defined by
$$(h*u)_n := \sum_{j \geq 0} h_j u_{n-j}, ~~~1 \leq n \leq m,$$ 
where it is assumed that $u_n := 0$ for $n\leq 0$.
Without any reference to a sampling or a reconstruction operator, noise shaping in this setting refers to the fact that the ``quantization noise'' $y-q$ is spectrally aligned with $h$. Note that the operator $H:u \mapsto h*u$ is invertible on $\R^m$ for any $m$, and therefore given any $y$ and $q$, there exists $u\in \R^m$ which satisfies \eqref{y-q-h-u}; this is trivial. However, the requirement that $\|u\|_\infty$ must be controlled uniformly in $m$ imposes restrictions on what $q$ can be for a given $y$; these solutions are certainly non-trivial to find and may not always exist.

The operator $H$ above (defined as convolution by $h$) is a lower triangular Toeplitz matrix with unit diagonal. With this view, let us relax the notion of a noise-shaping quantizer and assume that $H$ is any lower triangular $m \times m$ matrix with unit diagonal. We will refer to $H$ as a noise-shaping transfer operator where the associated noise-shaping relation is given by
\begin{equation} \label{y-q-H-u}
  y - q = Hu.
\end{equation}

Suppose we are given a sequence $(H_m)_1^\infty$ of $m \times m$ noise-shaping transfer operators. In this general setting, we say that an associated sequence $(Q_m)_1^\infty$ of quantizer maps (for which $q:=Q_m(y)$ and $u$ is determined by \eqref{y-q-H-u}) achieves noise shaping for $(H_m)$, $J$, and $\sA$, if $\|u\|_\infty \leq C$ for some constant $C$ independent of $m$. A slightly weaker assumption is to only require that $\|u\|_\infty = o(\|H_m^{-1}\|_{\infty \to \infty})$, though we shall not need to work in this generality in this paper.

In many applications, one works with $(H_m)_1^\infty$ which are ``progressive'' (also called ``nested'') in the sense that 
$$P_m \circ H_{m+1} \circ P_{m+1} = H_m \circ P_m,$$ 
where $P_m$ is the restriction of a vector to its first $m$ coordinates. Convolution is a standard example. In this case, it may be natural to require that the $(Q_m)_1^\infty$ are progressive as well. The classical $\sd$ modulation we saw in Section \ref{classical} is of this type. However, our general formulation does not impose progressiveness. 

As indicated earlier, noise-shaping quantizers provide non-trivial solutions to \eqref{y-q-H-u} and therefore do not exist unconditionally, though under certain suitable assumptions on $H$, $J$, and $\sA$, they exist and can be implemented via recursive algorithms. The simplest is the (non-overloading) {\em greedy quantizer} whose general formulation is given below:

\begin{proposition}\label{P:greedy}
Let $\sA := \sA_{L,\delta}$ denote the arithmetic progression in $\R$ which is of length $L$, spacing $2\delta$, and symmetric about $0$.
Assume that $H = I - \tilde H$, where $\tilde H$ is strictly lower triangular, and $\mu \geq 0$ such that 
$\|\tilde H\|_{\infty \to \infty} + \mu/\delta \leq L$. Suppose
$\|y\|_\infty \leq \mu$. For each $n\geq 1$, let
$$ q_n := \mathrm{round}_\sA\left (y_n + \sum_{j=1}^{n-1} \tilde H_{n,n-j} u_{n-j} \right)$$
and
$$ u_n := y_n + \sum_{j=1}^{n-1} \tilde H_{n,n-j} u_{n-j} -q_n .$$
Then the resulting $q$ satisfies \eqref{y-q-H-u}
with $\|u\|_\infty \leq \delta$.
\end{proposition}

This quantizer is called greedy because for all $n$, the selection of $q_n$ over $\sA$ is made so as to minimize $|u_n|$. The proof of this basic result follows easily by induction once we note that for any $w \in [-L\delta,L\delta]$, we have $|w - \mathrm{round}_\sA(w)| \leq \delta$, hence the scalar quantizer $\mathrm{round}_\sA$ is not overloaded. For details, see \cite{CG14}. Note that the greedy quantizer is progressive if $(H_m)_1^\infty$ is a progressive sequence of noise-shaping transfer operators. In the special case $Hu = h*u$ where $h_0 = 1$, we simply have $\|\tilde H\|_{\infty \to \infty} = \|h\|_1 - 1$. This special case is well-known and widely utilized (e.g. \cite{Candy-Temes, NST96, SchTe04, exp_decay}).

\subsection{Canonical duals of frames for noise shaping}\label{canonical-duals}

The earliest works on noise-shaping quantization in the context of finite frames used $\sd$ quantization and focused on canonical duals for reconstruction. Before we begin our discussion of these contributions we remind the reader of our convention: we identify an analysis frame with (the rows of) its analysis operator and a synthesis frame with (the columns of) its synthesis operator.

Let $\Phi$ be a finite frame and $y=\Phi x$ be the frame measurements of a given signal $x$. Assume that we quantize $y$ using a noise-shaping quantizer with transfer operator $H$.
Any left-inverse (dual) $\Psi$ of $\Phi$ gives
\begin{equation}\label{error_PsiHu}
x - \Psi q = \Psi(y - q) = \Psi Hu. 
\end{equation}

Using this expression, and specializing to the case of first order $\sd$ quantization, i.e., $H=D$ where $D$ is the lower bidiagonal matrix whose diagonal entries are 1 and subdiagonal entries are -1, \cite{BPY} observed that the reconstruction error can be bounded as 
\begin{equation}\label{eq:err_bound_BPY0}
\|x - \Psi q\|_2 \leq \|u\|_\infty \sum_{j=1}^m \|(\Psi D)_j\|_2
\end{equation}
where $(\Psi D)_j$ denotes the $j$th column of $\Psi D$. This led \cite{BPY} to introduce the notion of frame variation
\begin{equation}\label{eq:frame_var}
\mathrm{Var}(\Psi):=\sum_{j=1}^m \|\psi_j -\psi_{j+1}\|_2
\end{equation}
with $\psi_j$ denoting the $j$th column of $\Psi$ and $\psi_{m+1}$ defined to be zero. Using normalized tight-frames, i.e., frames $\Phi$ for which $\Phi^*\Phi = (m/k)I$, this resulted in the error bound
\begin{equation} \|x - \Phi^\dagger q\|_2 \leq  \frac{k}{m}  \|u\|_\infty \mathrm{Var}(\Phi^*), \label{eq:err_bound_BPY}\end{equation}
where $\Psi = \Phi^\dagger$ denotes the {\em canonical dual} of $\Phi$ defined (for an arbitrary frame $\Phi$) by
\begin{equation}\label{Phi_dagger}
\Phi^\dagger := (\Phi^*\Phi)^{-1}\Phi^*.
\end{equation}
 Subsequently, similarly defined higher-order frame variations were used to study the behavior of higher-order $\sd$ schemes (e.g., in \cite{BPY2} and \cite{BPA2007}) with corresponding generalizations of \eqref{eq:err_bound_BPY} and the conclusion that frames with lower variations lead to better error bounds. This motivated considering frames obtained via uniform sampling of smooth curves in $\R^k$ (called {\em frame paths}). As it turned out, however, this type of analysis based on frame-variation bounds does not provide higher-order reconstruction accuracy unless the frame path terminates smoothly. Smooth termination of the frame path is not available for most of the commonly encountered frames, and finding frames with this property can be challenging. Indeed, designing such frames was a main contribution of \cite{BPA2007} which showed a reconstruction error bound decaying as $m^r$ for $r$th order $\sd$ quantization of measurements using these frames. 
 
In practice, however, one must often work with a given frame rather than design a frame of their choosing. In such cases there are frames, sampled from smooth curves, for which reconstructing with the canonical dual yields reconstruction error that is \emph{lower bounded} by a term behaving like $m^{-1}$, regardless of the $\sd$ scheme's order $r\geq3$    (see, \cite{LPY} for the details). Consequently, to achieve better error decay rates one must seek either different quantization or different reconstruction schemes. We will consider both routes to improving the error bounds in what follows.

\subsection{Alternative duals of frames for noise shaping} \label{alternative-duals}

The discussion in Section \ref{canonical-duals} was based on canonical duals and it involved a particular method to bound the $2$-norm of the reconstruction error $x-\Psi q$, assuming $u$ is bounded in the $\infty$-norm. It is possible to significantly improve the reconstruction accuracy by allowing for more general duals, here called {\em alternative duals}. To explain this route, we return to the general noise-shaping quantization relation  \eqref{error_PsiHu}. We assume again that $u$ is known to be bounded in the $\infty$-norm, which is essentially the only type of bound available.
Hence, the most natural reconstruction error bound is given by 
\begin{equation}\label{error_bound_PsiHu}
 \|x - \Psi q \|_2 \leq \|\Psi H\|_{\infty \to 2} \|u \|_\infty.
\end{equation}

With this bound, the natural objective would be to employ an alternative dual $\Psi$ of $\Phi$ which minimizes $\|\Psi H\|_{\infty \to 2}$. An explicit solution for this problem is not readily available mainly because there is no easily computable expression for $\|A\|_{\infty \to 2}$ for a general $k \times m$ matrix $A$, so we replace it by a simpler upper bound. In fact, this was already done in \eqref{eq:err_bound_BPY0} because we have
\begin{equation}\label{bound_inf_2_1}
\|A\|_{\infty \to 2} \leq \sum_{j=1}^m \|A_j\|_2
\end{equation}
where again $A_j$ denotes the $j$th column of $A$. (This upper bound is also known to be the $L_{2,1}$-norm of $A$.) Another such bound which is often (but not always) better is given by
\begin{equation}\label{bound_inf_2_2}
\|A\|_{\infty \to 2} \leq \sqrt{m} \|A\|_{2 \to 2}.
\end{equation}
(Indeed, for a large random matrix with standard Gaussian entries, the upper bound in \eqref{bound_inf_2_2} behaves as $m + \sqrt{mk}$ whereas that of \eqref{bound_inf_2_1} behaves as $m\sqrt{k}$. Both of these upper bounds are easily seen to be less than $\sqrt{m} \|A\|_\mathrm{Fr}$, however.)

With this upper bound, we minimize $\|\Psi H\|_\mathrm{2 \to 2}$ over all alternative duals $\Psi$ of $\Phi$. Then an explicit solution is available and is given by 
\begin{equation}\label{opt-Psi-H}
\Psi_{H^{-1}} := (H^{-1}\Phi)^\dagger H^{-1}. 
\end{equation}
This idea was initially introduced specifically for $\sd$ quantization \cite{LPY,BLAY} with the choice $H = D^r$. The resulting alternative duals were called {\em Sobolev duals} and will be discussed in the next subsection. The above generalized version was stated in \cite{GLPSY} where the notation $\Psi_H$ and the term ``$H$-dual'' were introduced for the right hand side of \eqref{opt-Psi-H}, but because of a further generalization we will discuss in Section \ref{V-duals}, we find it more appropriate to use the label $H^{-1}$.

Note that the no noise-shaping case of $H=I$ yields the canonical dual. In general, 
we have 
$$\|\Psi_{H^{-1}} H\|_{2\to 2} = \|(H^{-1}\Phi)^\dagger\|_{2\to 2}
=\frac{1}{\sigmin(H^{-1}\Phi)}$$ 
so that \eqref{error_bound_PsiHu} and \eqref{bound_inf_2_2} yield the error bound
\begin{equation}\label{error_PsiHu_final}
  \|x - \Psi_{H^{-1}} q \|_2 
  \leq \frac{\sqrt{m}}{\sigmin(H^{-1}\Phi)}\|u \|_\infty.
\end{equation}

\subsubsection{Sobolev Duals}

In the case of $\sd$ modulation, $H$ is defined by \eqref{sig-del}, and given in matrix form by $D^r$ where the diagonal entries of the lower bidiagonal matrix $D$ are $1$ and the subdiagonal entries are $-1$. Because $\|\Psi D^r \|_{2 \to 2}$ resembles a Sobolev norm on $\Psi$, the corresponding alternative dual was called the ($r$th order) Sobolev dual of $\Phi$ in \cite{BLAY}. In this work, Sobolev duals of certain deterministic frames, such as the harmonic frames, were studied. More precisely, \cite{BLAY} considered frames obtained using a sufficiently dense sampling of vector-valued functions on $[0,1],$ which had the additional property that their component functions were piecewise $C^1$ and linearly independent. For such frames, it was shown that 
\begin{equation}\label{sigmin_smooth}
\sigma_\mathrm{min} (D^{-r} \Phi) \geq c_r m^{r + \frac{1}{2}},
\end{equation}
hence with \eqref{error_PsiHu_final}, the reconstruction error using the $r$th order Sobolev dual satisfies 
\begin{equation}
\|x-\Psi_{D^{-r}}q\|_2 \leq \frac{C_{r}}{ m^{r}} \|u\|_\infty
\label{eq:BLAY_error}
\end{equation}
with $C_r := 1/c_r$.
Here, for a fixed stable $\sd$ scheme, the constant $C_r$ depends only on the order $r$ and the vector-valued function from which the frame was sampled.
The main technique used in \cite{BLAY} to control the operator norm $\|\Psi_{D^{-r}} D^r \|_{2 \to 2}$ is a Riemann sum argument. The argument leverages the smoothness of the vector-valued functions from which the frames are sampled to obtain a lower bound on $\|D^{-r}\Phi x\|_2$ over unit norm vectors $x\in \R^d$ and produces the stated lower bound \eqref{sigmin_smooth}.

As mentioned before, error bounds similar to \eqref{eq:BLAY_error} had also been obtained in \cite{BPA2007}, albeit for specific tight frames. 
Nevertheless, in both \cite{BLAY} and \cite{BPA2007}, the decay of the error associated with $\sd$ quantization is a polynomial function of the number of measurements. The significance of this polynomial error decay stems from the fact that for any frame, a lower bound on the reconstruction error associated with MSQ is known to decay only linearly in $m$ \cite{GVT98}.  

\subsubsection{Refined Bounds Using Sobolev Duals}
The analysis of \cite{BLAY} was refined in \cite{KSW} in two special cases: harmonic frames, and the so-called Sobolev self-dual frames. For these frames, \cite{KSW} established an upper bound on the reconstruction error that decays as a root-exponential function of the number of measurements. More specifically, for harmonic frames, \cite{KSW} explicitly bounds the constant $C_r$ in \eqref{eq:BLAY_error} and, as in \cite{exp_decay} and \cite{DGK}, optimizes the $\sd$ scheme's order $r$ as a function of the number of measurements. Quantizing with a $\sd$ scheme of the optimal order $r_\mathrm{opt}(m)$ and reconstructing with the associated Sobolev dual results in a root-exponential error bound
\begin{equation}\|x-\Psi_{D^{-r_\mathrm{opt}}}q\|_2 \leq c_1 e^{-c_2\sqrt{m/k}}
\label{eq:KSW_error1}\end{equation}
where the constants $c_1$ and $c_2$ depend on the quantization alphabet $\sA_{L,\delta}$ and possibly on $k$ as well. This possible dependence on $k$ is absent in the similar bound for Sobolev self-dual frames. Sobolev self-dual frames are defined using the singular value decomposition $D^{r} = U\Sigma V^*$.  Here, the $m\times k$ matrix corresponding to a Sobolev self-dual frame consists of the $k$ columns of $U$ associated with the smallest singular values of $D^r$. This construction implies that the frame admits itself as both a canonical dual and Sobolev dual of order $r$, hence the name. More importantly, this construction also allows one to bound $C_r$ in \eqref{eq:BLAY_error} explicitly and optimize the $\sd$ scheme's order $r$ to obtain the error bound \eqref{eq:KSW_error1}, without any dependence of the constants on $k$. 

While we have so far discussed deterministic constructions of frames, Gaussian random frames were studied in \cite{GLPSY}, and later, sub-Gaussian random frames in \cite{subGaussian}. We will discuss these random frames extensively in Section \ref{random-Sobolev}, though at this point we note that, like the harmonic and Sobolev self-dual frames, these frames also allow for root-exponential error decay when the order of the $\sd$ scheme is optimized.  

In the context of $\sd$ quantization of frame coefficients using a fixed alphabet $\sA$, the number of measurements is proportional to the total number of bits. Hence, the error bounds \eqref{eq:BLAY_error} and \eqref{eq:KSW_error1} can be interpreted as polynomially and root-exponentially decaying in the total number of bits. While these bounds are certainly a big improvement over the linearly decaying lower bound associated with MSQ, they are still sub-optimal. To see this, one observes that the problem of quantizing vectors in the unit ball of $\R^k$ with a maximum reconstruction error of $\epsilon$ is analogous to covering the unit-ball with balls of radius $\epsilon$.  A simple volume argument shows that to quantize the unit ball of $\R^k$ with an error of $\varepsilon$, one needs at least $k\log_2 \big(\frac{1}{\epsilon}\big)$ bits. Thus, the reconstruction error can at best decay exponentially in the number of bits used.  Moreover, since there exists a covering of the unit-ball with no more than $\big(
\frac{3}{\epsilon}\big)^k$ elements (see, e.g., \cite{lorentz1996constructive}), in principle an exponential decay in the error as a function of the number of bits used is possible. This exponential error decay is predicated on a quantization scheme that has direct access to $x$ and, more importantly, the ability to compare $x$ to each of the approximately $\epsilon^{-k}$ elements of the covering, to assign it an appropriate binary label. The reconstruction scheme for this quantization would then simply replace the binary label by the center of the element of the covering associated with it. Of course, this setting is markedly different from the noise-shaping quantization of frame coefficients considered in this chapter, but it establishes exponential error decay in the number of bits as optimal. 

To achieve exponential error decay in the number of bits, \cite{IS13} proposed an encoding scheme to follow $r$th order $\sd$ quantization.  The encoding scheme consists of  using an $\ell \times m$ Bernoulli random matrix $B$, with $\ell$ slightly larger than $k$,  to embed  the vector $D^{-r}q$ into a lower dimensional subspace. Since $B$ serves as a distance-preserving Johnson-Lindenstrauss embedding (see, \cite{JLoriginal, achlioptas2001}), the vector $BD^{-r}q$ effectively contains all the information needed for accurate reconstruction of $x$, and it is the only quantity retained. Moreover, the number of bits required to store $BD^{-r}q$ scales only logarithmically in $m$. Using $(BD^{-r}\Phi)^\dagger$ as a reconstruction operator (acting on $BD^{-r} q$) and employing the properties of Johnson-Lindenstrauss embeddings, \cite{IS13} shows that the reconstruction error still decays as it would have if no embedding had been employed. In particular, this means an error decay of $m^{-r}$ for the frames 
discussed in this section. Combining these two observations, i.e., logarithmic scaling of the number of bits with $m$, and polynomial decay of the error, \cite{IS13} obtains reconstruction error bounds that decay \emph{exponentially}, i.e., near optimally, in the number of bits.

It turns out that exponential decay of the reconstruction error (in the bit rate or in the oversampling ratio $m/k$) can also be achieved by means of the ``plain route'' of noise-shaping quantization and alternative dual reconstruction only, but with noise-shaping unlike $\sd$ quantization and more like the conventional beta encoding \cite{Chou_thesis, CG14}. This method, called beta duals, is explained next for general frames, and later in Section \ref{random-beta} for random frames. 

\subsubsection{Further generalizations: $V$-duals}
\label{V-duals}

Given any $m \times k$ matrix $\Phi$ whose rows are a frame for $\R^k$, consider any $p \times m$ matrix $V$ (i.e., not necessarily square) such that $V\Phi$ is also a frame for $\R^k$. We will call 
\begin{equation}\label{Phi:vdual}
\Psi_{\,V} := (V\Phi)^\dagger V
\end{equation}
the {\em $V$-dual} of $\Phi$. (The square and invertible case of $V = H^{-1}$ was already discussed at the beginning of this subsection.) When $p < m$, we call $V\Phi$ the $V$-{\em condensation} of $\Phi$.

With a $V$-dual, we have $\Psi_{\,V} H = (V\Phi)^\dagger VH$ so that 
\begin{equation}\label{error_PsiVu}
\|\Psi_{\,V} H\|_{\infty \to 2} 
\leq  \frac{\| VH\|_{\infty \to 2} }{\sigmin( V\Phi) }
\leq \frac{\sqrt{p} \| VH\|_{\infty \to \infty} }{\sigmin( V\Phi) }.  
\end{equation}

For $V=H^{-1}$ (and therefore, $p=m$), combination of \eqref{error_bound_PsiHu} with 
\eqref{error_PsiVu} agrees with \eqref{error_PsiHu_final}. However, as shown in \cite{CG14}, optimization of \eqref{error_PsiVu} over $V$ can produce a strictly smaller reconstruction error upper bound. A highly effective special case is discussed next.

\paragraph{\bf Beta duals}
Beta duals have been recently proposed and studied in \cite{Chou_thesis, CG14}. They
constitute a special case of $V$-duals, while they relate strongly to classical beta expansions. (See \cite{P, DK} for the classical theory of beta expansions, and \cite{DDGV} for the use of beta expansions in A/D conversion as a robust alternative to successive approximation.)
In order to illustrate the main construction of beta duals without technical details, our presentation in this article will be restricted to certain dimensional constraints as described below. 

Let $m \geq p \geq k$ and assume that $\lambda' := m/p$ is an integer.
For any $\beta > 1$, let $h^\beta$ be the (length-$2$)
sequence given by $h^\beta_0 = 1$ and $h^\beta_1 = -\beta$.
Define $H^\beta$ to be the $\lambda' \times \lambda'$ noise-shaping transfer operator corresponding to $h^\beta$, and 
$$v^\beta := [\beta^{-1} ~~ \beta^{-2}~\cdots~~\beta^{-\lambda'}].$$
We set 
\begin{equation}\label{H-V-beta}
H:=\left[\begin{matrix} H^\beta  & & \\ & \ddots & \\ & & H^\beta \end{matrix}\right]_{m \times m}
\quad\text{and}\quad
V:=\left[\begin{matrix} ~~~v^\beta~~~  & & \\ & ~~~\ddots~~~ & \\ & & ~~~v^\beta~~~ \end{matrix}\right]_{p \times m}.
\end{equation}
In other words, $H = I_p \otimes H^\beta$ and $V=I_p \otimes v^\beta$ where $\otimes$ denotes the Kronecker product. It follows that $VH = I_p \otimes (v^\beta H^\beta)$.
Since $v^\beta H^\beta = [0~\cdots~0~~\beta^{-\lambda'}]$, we have 
$\|VH \|_{\infty\to\infty} = \beta^{-\lambda'}$ which, together with \eqref{error_bound_PsiHu} and \eqref{error_PsiVu}, yields
 
\begin{equation} \label{betadual_bound}
\|x - \Psi_{V} q\|_2 \leq 
\frac{\sqrt{p} \|u\|_\infty}{\sigmin(V\Phi)} \beta^{-\lambda'}.
\end{equation} 

For certain special frames, such as the harmonic semi-circle frames, it is possible to set $p$ as low as $k$ and turn the above bound into a near-optimal one in terms of its bit-rate \cite{CG14}. 
The case of random frames will be discussed in the next section. 

\begin{figure}[htp]
\centering
\centerline{\includegraphics[scale=0.75]{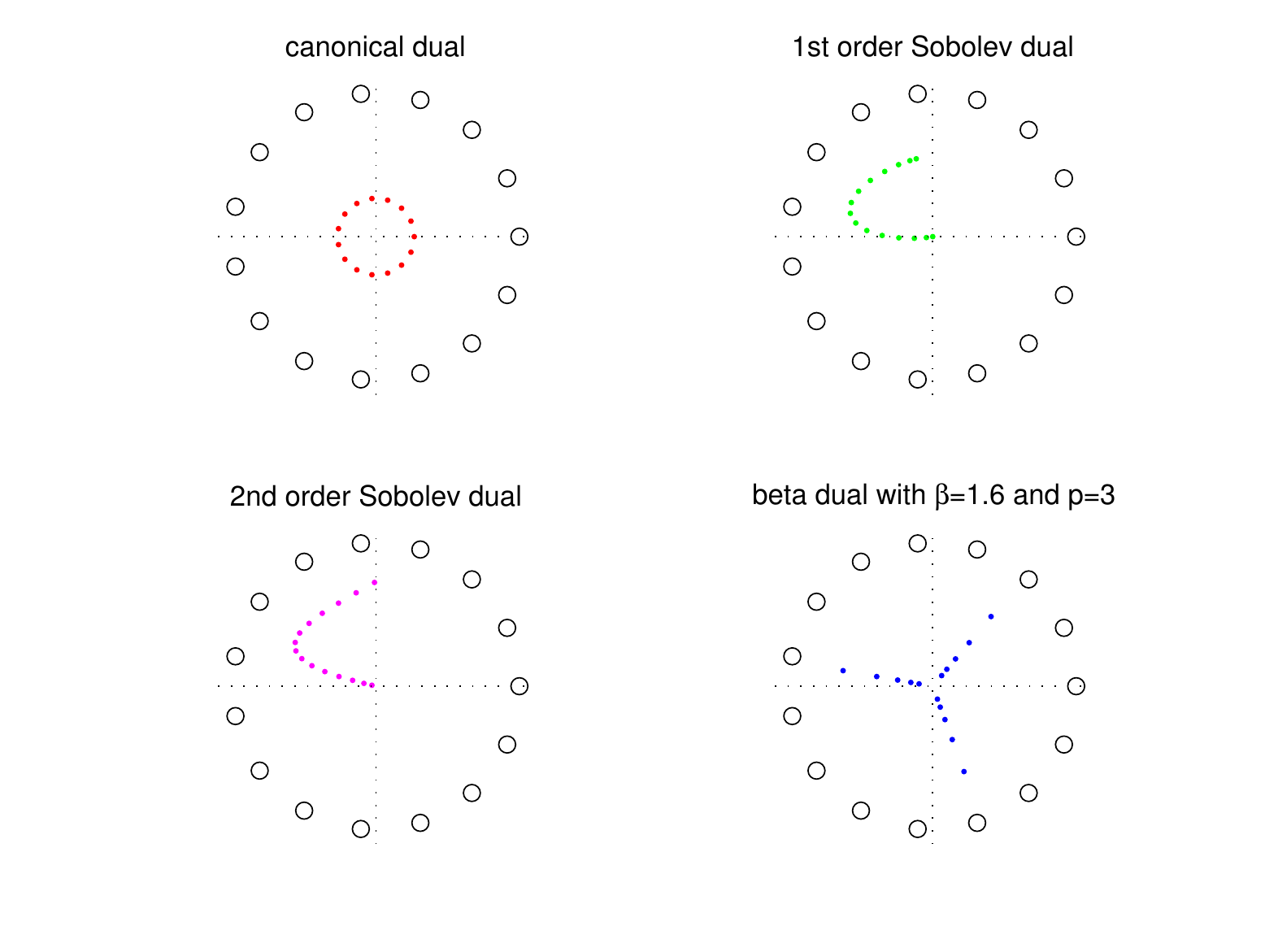}}
\caption{Comparative illustration of the various alternative duals described in this paper:
Each plot depicts the original frame in $\R^2$ consisting of the $15th$ roots-of-unity along with one of its duals (scaled up by a factor of two for visual clarity). For the computation of the alternative duals, the analysis frame was ordered counter-clockwise starting from $(1,0)$.}
\label{fig:4_duals_hf}
\end{figure}

In Fig. \ref{fig:4_duals_hf}, we illustrate a beta dual of a certain ``roots-of-unity'' frame along with the Sobolev duals of order 0 (the canonical dual), 1, and 2.

\medskip

\section{Analysis of Alternative Duals for Random Frames}
\label{random-frames}
In this section, we consider random frames, that is, frames whose analysis (or synthesis) operator is a random matrix. Certain classes of random matrices have become of considerable importance in high dimensional signal processing, particularly with the advent of compressed sensing. One main reason for this is that their inherent independence entails good conditioning of not only the matrix, but also its submatrices. Because of the fast growing number of such submatrices with dimension, the latter is very difficult to achieve with deterministic constructions. This also means, however, that any two frame vectors are approximately orthogonal, so frame path conditions that would imply recovery guarantees using canonical dual frames will almost never hold. For this reason, it is crucial to work with alternative duals. We separately consider the two main examples discussed above, Sobolev duals and beta duals.

\subsection{Sobolev duals of random frames}
\label{random-Sobolev}
As noted above, the Sobolev dual of a frame is the dual frame $\Psi$ that minimizes the expression $\|\Psi D^r\|_{2\rightarrow 2}$, and the explicit minimizer is given by \eqref{opt-Psi-H} with $H=D^{r}$. By \eqref{error_PsiHu_final}, a bound for the error that arises when using this alternative dual to reconstruct is governed by $\sigmin(D^{-r}\Phi)$. Thus a main goal of this subsection is to discuss the behavior of this minimum singular value.
 
The matrix $D^{-r}\Phi$ is the product of a deterministic matrix $D^{-r}$, whose singular values are known to a sufficient approximation, and a random matrix $\Phi$, whose singular values are known to be well concentrated. Nevertheless, using a product bound does not yield good results, mainly because the singular values of $D^{-r}$ differ tremendously, so any worst case bound will not be good enough. One approach to provide a refined bound is to first provide lower bounds for the action of $D^{-r}\Phi$ on a single vector and then proceed via a covering argument. That is, one combines these lower bounds for all of the vectors forming an $\epsilon$-net, obtaining a uniform bound for the net. An approximation argument then allows to pass from the net to all vectors in the sphere. In this way, \cite{GLPSY} obtains the following result for Gaussian random frames:

\begin{theorem}[\cite{GLPSY}] \label{Sobolev_dual_bound}
Let $\Phi$ be an $m\times k$ random matrix whose entries are
i.i.d.~standard Gaussian variables. Given $r \in \mathbb{N}$ and $\alpha \in (0,1)$, there exist constants strictly positive $r$-dependent constants $c_1$, $c_2$, and $c_3$ such that
if $\lambda:=m/k \geq  (c_1 \log m)^{1/(1-\alpha)}$, then with probability at
least $1 - \exp(-c_2 m \lambda^{-\alpha})$,
\begin{equation}\label{GLPSY-sigmin}
\sigma_{\min}(D^{-r}\Phi) \geq c_3(r) \lambda^{\alpha(r-\frac{1}{2})}\sqrt{m}.
\end{equation}
\end{theorem}

In this approach, one explicitly uses the density of the Gaussian distribution. Thus, as soon as the matrix entries fail to be exactly Gaussian, a completely different approach is needed. In what follows, we will present the main idea of the method used in \cite{subGaussian} to tackle the case of random matrices with independent sub-Gaussian entries as introduced in the following definition (for alternative characterizations of sub-Gaussian random variables see, for example, \cite{ve12-1}). This approach is also related to the RIP-based analysis for quantized compressive sampling presented in \cite{FK14} (cf.~Section~\ref{CS} below).

\begin{definition}\label{def:subgvar} A random variable $\xi$ is sub-Gaussian with parameter $c>0$ if it satisfies ${\mathbb P}(|\xi| >t) \leq e^{1-ct^2}$ for all $t\geq 0$.
\end{definition}

As in the Gaussian case presented in \cite{GLPSY}, we employ the singular value decomposition $D^{-r}=U \Sigma V^*$ where $U$ and  $V$ are unitary and $\Sigma\in \R^{m\times m}$ is a diagonal matrix with entries $s_1\geq  \dots \geq s_m \geq 0$. Then
\begin{equation*}
 \sigmin(D^{-r}\Phi)  = \sigmin (U \Sigma V^* \Phi) = \sigmin ( \Sigma V^* \Phi),
\end{equation*}
as $U$ is unitary. Furthermore, for $P_\ell:\R^m\rightarrow \R^\ell$ the projection onto the first $\ell$ entries, $\ell\leq m$, one has in the positive semidefinite partial ordering $\succeq$
\begin{equation*}
\Sigma \succeq P_\ell \Sigma = P_\ell \Sigma P_\ell^* P_\ell \succeq s_\ell P_\ell.
\end{equation*}
Here the first inequality uses that $P_\ell$ is a projection, the following equality uses that $\Sigma$ is diagonal, and the last inequality uses that the diagonal entries of $\Sigma$ are ordered.

As a consequence, we find that $\sigmin(D^{-r}\Phi) \geq s_\ell \sigmin(V^*\Phi)$. For Gaussian matrix entries, this immediately yields Theorem~\ref{Sobolev_dual_bound}, as standard Gaussian vectors are rotation invariant, so $P_\ell V^*\Phi$ is just a standard Gaussian matrix, whose singular value distributions are well understood (see for example \cite{ve12-1}). Applying the bound for different values of $\ell$ yield the theorem for different choices of $\alpha$.

For independent, zero mean, unit variance sub-Gaussian (rather than Gaussian) matrix entries, one no longer has such a strong version of rotation invariance; while the columns of $V^* \Phi$ will still be sub-Gaussian random vectors, its entries will, in general, no longer be independent.  There are also singular value estimates that require only independent sub-Gaussian matrix columns rather than independent entries (see again \cite{ve12-1}), but such bounds require that the matrix columns are of constant norm. Even if $\Phi$ and hence also $V^*\Phi$ has constant norm columns (such as for example for Bernoulli matrices, $\Phi_{ij}\in \pm 1$), the projection $P_\ell$ will typically map them to vectors of different length.

In order to nevertheless bound the singular values, we again use a union bound argument, first considering the action on one fixed vector $x$ of unit norm. Then we write
\[
 \|V^*\Phi x\|_2^2 = \sum_{i, i'=1}^k \sum_{j,j'=1}^m  x_i  \Phi_{j i} (V P_\ell^* P_\ell V^*)_{j j'} \Phi_{j' i'} x_{i'}.
\]
Thus $\|V^*\Phi x\|_2^2$ is a so-called chaos process, that is, a random quadratic form of the form $\langle \xi, M \xi\rangle$, where $\xi$ is a random vector with independent entries (in this case, the vectorization of $\Phi$). Its expectation is given by 
\[
{\mathbb E} \|V^*\Phi x\|_2^2 = \sum_{i=1}^k \sum_{j=1}^m  x_i^2  {\mathbb E} \Phi_{j i}^2 (V P_\ell^* P_\ell V^*)_{j j} = \|x\|_2^2 {\operatorname{tr}}(V P_\ell^* P_\ell V^*) =\ell,
\]
where the last equality uses the cyclicity of the trace. Its deviation from the expectation can be estimated using the following refined version of the Hanson-Wright inequality, which has been provided in \cite{rv13} (see \cite{HW71} for the original version).

\begin{theorem}\label{thm:HW}
 Let $\xi=(\xi_1,\dots, \xi_n)\in\R^n$ be a random vector with independent components $\xi_i$ which are sub-Gaussian with parameter $c$ and satisfy ${\mathbb E}  \xi_i=0$. Let $A$ be an $n\times n$ matrix. Then for every $t\geq 0$, 
 \[
  {\mathbb P}\big\{|\langle \xi, M \xi\rangle - {\mathbb E} \langle \xi, M \xi\rangle|>t\big\} \leq 2\exp\Big(-C_4 \min\big(\frac{t^2}{c^4 \|M\|_{F}^2}, \frac{t}{c^2\|M\|_{2\rightarrow 2}}\big)\Big),
 \]
where $C_4$ is an absolute constant.
\end{theorem}

To obtain a deviation bound for the above setup, we thus need to estimate the Frobenius norm $\|M\|^2_F :={\operatorname{tr}} M^*M= \sum_{i,i',j,j'} M_{(i,i'),(j,j')}^2$ and the operator norm $\|M\|_{2\rightarrow 2}:= \sup_{\|y\|_2=1} \|My\|_2$ of the doubly-indexed matrix $M$ given by $M_{(i,j),(i',j')}=x_i x_{i'} (V P_\ell^* P_\ell V^*)_{j j'} $. For the Frobenius norm, we write 
\[
 \|M\|^2_F = \sum_{i,i',j,j'} x_i^2 x_{i'}^2 (V P_\ell^* P_\ell V^*)_{j j'}^2 = \| V P_\ell^* P_\ell V^*\|_F^2 = {\operatorname{tr}} (V P_\ell^* P_\ell V^*V P_\ell^* P_\ell V^*) = \ell,
\]
where in the last equality, we used again the cyclicity of the trace,  that $V$ is unitary, and that $P_\ell^*P_\ell$ is a projection. For the operator norm, we note that\[
M=  P_{ \ell} V^* \left( \begin{matrix} x^T & 0 & \cdots & 0\\
0 & x^T & \cdots & 0\\
\vdots & \vdots & \vdots & \vdots\\
0 & \cdots & 0 & x^T
\end{matrix} \right),                                                                                                                                                                                                                                                                                                                                                 \]
so as all these three factors have operator norm $1$, the norm of their product is bounded above by $1$. On the other hand, applying $M$ to the unit norm vector $y$ given by $y_{(i,j)}= x_i V_{1j}$ yields $My=e_1$, where $e_1$ is the first standard basis vector,  showing that the norm is also lower bounded by $1$. So one indeed has $\|M\|_{2\rightarrow 2}=1$. Combining these bounds with Theorem~\ref{thm:HW} yields the following generalization of Theorem~\ref{Sobolev_dual_bound} for sub-Gaussian frames.

\begin{theorem}[\cite{subGaussian}]
\label{thm:main_1} 
Let $\Phi$ be an $m\times k$ random matrix whose entries are zero mean, unit variance, sub-Gaussian random variables with parameter $c$. 
Given $r \in \mathbb{N}$ and $\alpha \in (0,1)$, there exist constants $c=c(r)>0$ and $c'=c'(r)>0$ such that if $\lambda:= \frac{m}{k} \geq c^{\frac{1}{1-\alpha}}$ then one has with probability at
least $1 - \exp(-c' m \lambda^{-\alpha})$
\begin{equation}\label{KSY-sigmin}
\sigma_{\min}(D^{-r}\Phi) \geq \lambda^{\alpha(r-\frac{1}{2})}\sqrt{m}.
\end{equation}
\end{theorem}

Combining \eqref{error_PsiHu_final} for $H=D^r$ with the lower bound of \eqref{GLPSY-sigmin} or \eqref{KSY-sigmin}, the Sobolev dual reconstruction $\Psi_{D^{-r}}q$ from $\sd$ quantized frame coefficients $y=\Phi x$ results in the error bound 
\begin{equation} \label{Sobolev-random-err-bound}
\|x - \Psi_{D^{-r}}q \|_2 \leq C(r)  \lambda^{-\alpha(r-\frac{1}{2})} \| u\|_\infty.
\end{equation}

Thus the error decays polynomially in the oversampling rate $\lambda$ as long as the underlying $\sd$ scheme is stable. For the greedy quantization rule, stability follows from Proposition \ref{P:greedy}, as long as $\|y\|_\infty \leq \mu$ for a suitable $\mu$ whose range is constrained by the quantization alphabet $\sA_{L,\delta}$ and $r$. (It can be easily computed that for $H=D^r$, we have $\|\tilde H\|_{\infty \to \infty} = 2^r - 1$. Hence we require $L > 2^r - 1$, with the value of $\delta$ assumed to be adjustable.) If we assume that $\|x\|_2 \leq 1$, then controlling $\|y\|_\infty$ amounts to bounding $\|\Phi\|_{2\to \infty}\leq \|\Phi\|_{2\to 2}$ and thus to bounding the maximum singular value of a rectangular matrix with independent sub-Gaussian entries. This is  a well-understood setup, it is known that the singular values of such a matrix are well concentrated and one has $\|\Phi\|_{2\to \infty}\leq \|\Phi\|_{2\to 2} = {O}(\sqrt m)$ with high probability (see again \cite{ve12-1}). As a consequence,
 the 
$\sd$ scheme is stable provided $L$ is chosen large enough and the quantizer level is adjusted accordingly. We conclude that sub-Gaussian frame expansions quantized using a greedy $r$-th order $\sd$ scheme allow for reconstruction error bounds decaying polynomially in the oversampling rate, where the decay order can be made arbitrarily large by choosing $r$ large enough.

\subsection{Beta duals of random frames}
\label{random-beta}

We return to the Gaussian distribution for the analysis of beta duals for random frames.
Based on the error bound \eqref{betadual_bound} derived in Section \ref{V-duals}, it now suffices to give a probabilistic lower bound for $\sigmin(V\Phi)$. Note that the entries of the $p\times k$ matrix $V\Phi$ are i.i.d.~Gaussian with variance 
\begin{equation}\label{sig-lambda}
\sigma^2_{\lambda'}:=\beta^{-2}+\cdots+\beta^{-2\lambda'}.
\end{equation}

At this point, a choice for the parameter $p$ needs to be made. In \cite{CG14}, both choices of $p=k$ and $p > k$ were studied in detail. The analysis of the former choice is somewhat cleaner, but the strongest probabilistic estimates follow by choosing $p$ greater than $k$.

We will primarily be interested in the smallest singular value of $V\Phi$ being near zero. For $p=k$, the following well-known result suffices:

\begin{theorem}[{\cite[Theorem 3.1]{rudelson2010non}}, {\cite{edelman1988eigenvalues}}]\label{P:squaregaussminsv}
Let $\Omega$ be a $k \times k$ random matrix with entries drawn
independently from $\mathcal{N}(0,\sigma^2)$. Then for any $\varepsilon > 0$,
\[ \mathbb{P}\left(\left\{ \sigmin(\Omega) \leq 
\varepsilon\sigma/\sqrt{k}\right\}\right) \leq \varepsilon. \]
\end{theorem}

Meanwhile, the stability of the greedy quantizer with alphabet $\sA_{L,\delta}$ can be ensured in a way similar to the case of Sobolev duals, noting that $\|\tilde H\|_{\infty \to \infty} = \beta$. Hence, we know that if $\beta + \mu/\delta \leq L$, then $\|u\|_\infty \leq \delta$. By standard Gaussian concentration results, $\mu \leq  4 \sqrt{m}$ is guaranteed with probability at least $1-e^{-2m}$. 
Therefore, with \eqref{betadual_bound} and Theorem \ref{P:squaregaussminsv} in which we set $\Omega = V\Phi$, we obtain
\begin{equation} \label{betadual_bound2}
\|x - \Psi_{V} q\|_2 \leq 
 kL \varepsilon^{-1} \delta \beta^{-m/k}
\end{equation} 
with probability at least $1-\varepsilon - e^{-2m}$, where we have also used the simple chain of inequalities $1/\sigma_{\lambda'} \leq \beta \leq L$. The value of $\beta$ can be chosen arbitrarily close to $L$ with sufficiently large values of $\delta$. However, the optimal choice would result from minimizing $\delta \beta^{-m/k}$ subject to $\beta + \mu/\delta = L$. For details, see \cite{CG14}.

For $p>k$, we have the following result:
\begin{theorem}[{\cite[Theorem 4.3]{CG14}}]\label{T:svmingaussrect}
Let $p > k$ and $\Omega$ be a $p \times k$ random matrix whose entries are drawn independently from $\mathcal{N}(0,\sigma^2)$. Then for any $0 < \varepsilon < 1$,
\[ \mathbb{P}\left(\left\{ \sigmin(\Omega) \leq \varepsilon\sigma \sqrt{p}/2\right\}\right) \leq
\left (10+8\sqrt{\log \varepsilon^{-1}}\right)^k  e^{p/2} \varepsilon^{p-k}.
 \]
\end{theorem}
The corresponding error bound 
\begin{equation} \label{betadual_bound3}
\|x - \Psi_{V} q\|_2 \leq 
 2 L \varepsilon^{-1} \delta \beta^{-m/p}
\end{equation}
now holds with higher probability. The choices $\varepsilon \approx \beta^{-\eta m/p}$ for small $\eta$ and $p \approx (1+\eta)k$ turn out to be good ones. For details, again see \cite{CG14}.

\section{Noise-shaping Quantization for Compressive Sampling}
\label{CS}

Compressive sampling (also called compressed sensing) has emerged over the last decade as a novel sampling paradigm. It is
based on the empirical observation that various important classes of
signals encountered in practice, such as audio and images, admit
(nearly) sparse approximations when expanded with respect to an
appropriate basis or frame, such as a wavelet basis or a Gabor
frame. Seminal papers by Cand{\`e}s, Romberg, and Tao \cite{CRT}, and
by Donoho \cite{Donoho2006} established the fundamental theory,
specifying how to collect the samples (or measurements), and the
relation between the approximation accuracy and the number of samples
acquired (``sampling rate'') vis-a-vis the sparsity level of the
signal. Since then the literature has matured considerably, again
focusing on the same issues, i.e., how to construct effective
measurement schemes and how one can control the approximation error as
a function of the sampling rate, e.g., see \cite{FoucartRauhutBook}.

By now compressive sampling is well-established as an
effective sampling theory. From the perspective of
practicability, however, it also needs to be accompanied by a
quantization theory. Here, as in the case of frames, MSQ is highly limited as a quantization strategy in terms of its rate-distortion performance. Thus, efficient quantization methods are needed for compressive sampling to live up to its name, i.e., to provide compressed representations in the sense of source coding. 

In this section, we will discuss how noise-shaping methods can be employed to
quantize compressive samples of sparse and compressible signals to vastly improve the reconstruction accuracy compared to the default method of MSQ. We start with the basic framework of compressive sampling as needed for our discussion.

\subsection{Basics of Compressive Sampling}\label{sec:CSintro}

In the basic theory of compressive sampling, the signals of interest are finite (but potentially high) dimensional vectors that are exactly or approximately
\emph{sparse}. More precisely, we say that a signal $x$ in $\R^N$ is
$k$-sparse if it is in $\Sigma_k^N:=\{ x\in \R^N:\ \|x\|_0\le
k\}$. Here $\|x\|_0$ denotes the number of non-zero entries of
$x$. The signals we encounter in practice are typically not sparse,
but they can be well-approximated by sparse signals. Such signals are
referred to as compressible signals and roughly identified as signals
$x$ with small $\sigma_k(x)_{\ell_p}$, \emph{the best $k$-term
  approximation error of $x$ in $\ell_p$}, defined by
$$
\sigma_k(x)_{\ell_p}:=\min_{z\in \Sigma_k^N} \|x-z\|_p.
$$

Compressive sampling consists of acquiring linear,
non-adaptive measurements of sparse or compressible signals, possibly
corrupted by noise, and recovering (an approximation to) the original
signal from the compressive samples via a computationally tractable
algorithm. In other words, the compressive samples are obtained by
multiplying the signal of interest by a \emph{compressive sampling (measurement) matrix}. The
success of recovery algorithms relies heavily on certain properties of
this matrix. To state this dependence precisely, we next define the
restricted isometry constants of a matrix.
\begin{definition} \label{def:RIP}
  The restricted isometry constant (see, e.g., \cite{CRT})
  $\gamma_k:=\gamma_k(\Phi)$ of a matrix $\Phi \in \R^{m\times N}$ is the smallest
  constant for which
$$(1-\gamma_k)\|x\|_2^2\leq \|\Phi x\|_2^2 \leq (1+\gamma_k )\|x\|_2^2
$$
for all $x\in \Sigma_k^N$.
\end{definition}

Suppose that $\Phi \in \R^{m\times N}$ is used as a compressive
sampling matrix. Here, $m$ denotes the number of measurements
and is significantly smaller than $N$, the ambient dimension of the
signal. Let $\tilde y:=\Phi x + w$ denote the (possibly) perturbed measurements of a signal $x \in \R^N$, where the unknown perturbation $w$ satisfies $\|w\|_2 \leq \epsilon$. A crucial result in the theory of compressive sampling states that if the restricted isometry constants of $\Phi$ are suitably controlled (e.g. as originally stated in \cite{CRT}, or more recently as in \cite{Cai_IEEE_2014} which only assumes $\gamma_{ak}\le \sqrt{(a-1)/a}$ for some $a\ge 4/3$), then there is an approximate recovery $\Delta_1^\epsilon(\Phi,\tilde y)$ of $x$ which satisfies
\begin{equation}\label{robrec}
\|x-\Delta_1^\epsilon(\Phi,\tilde y)\|_2\leq C \epsilon + D\sigma_k(x)_{\ell_1}/\sqrt{k}.
\end{equation} 
Here, $\Delta_1^\epsilon(\Phi,\tilde y)$ is found by mapping $\tilde y$ to a minimizer of a tractable, convex optimization problem---which is often called the ``Basis Pursuit Denoise'' algorithm---given by
$$\Delta_1^\epsilon(\Phi,\tilde y):=\arg\min_{z}\|z\|_1 \ \ \text{subject to} \ \
\|\Phi z - \tilde y\|_2\le \epsilon.
$$
$C$ and $D$ are constants that depend on $\Phi$, but can be made absolute by slightly stronger assumptions on $\Phi$.

Note that in the noiseless case, it follows from \eqref{robrec} that any $k$-sparse signal can be exactly recovered from its compressive samples as $\Delta_1^0(\Phi,\Phi x)$. In the general case, the approximation error remains within the noise level and within the best $k$-term approximation error of $x$ in $\ell_1$. Hence the recovery is robust with respect to the amount of noise and stable with respect to violation of the exact sparsity assumption. 
The decoder $\Delta_1^\epsilon$ is a \emph{robust compressive sampling decoder}  as defined next. 

\begin{definition}\cite[Definition 4.9]{subGaussian}
  Let $\varepsilon>0$, let $m,N$ be positive integers such that
  $m<N$ and suppose that $\Phi\in\R^{m\times N}$. We say that
  $\Delta: \R^{m\times N}\times\R^m\rightarrow \R^N$ is a robust compressive sampling decoder with
  parameters $(k, a,\gamma)$, $k<m$, and constant $C$ if
\begin{equation}\label{now26}
\|x-\Delta(\Phi, \Phi x+e)\| \le C \varepsilon, 
\end{equation}
for all $x\in \Sigma_k^N$, $\|e\|_2 \le \varepsilon$, and all matrices
$\Phi$ with a restricted isometry constant
$\gamma_{ak}<\gamma$. 
\end{definition}

Examples of robust decoders include $\Delta_1^\epsilon$ and its $p$-norm generalization $\Delta_p^\epsilon$ with $0<p\le1$
\cite{CRT,SY10}, compressive sampling matching pursuit (CoSaMP)
\cite{cosamp}, orthogonal matching pursuit (OMP) \cite{Zhang11}, and
iterative hard thresholding (IHT) \cite{Blumensath09}. See also
\cite{FoucartRauhutBook} for detailed estimates of the relevant
parameters.

\subsection{Noise-shaping Quantization of Compressive Samples}
Even though noise shaping methods are tailored mainly for quantizing
redundant representations, perhaps surprisingly, they also provide
efficient strategies for quantizing compressive samples
\cite{CISS10,GLPSY,subGaussian, FK14}. The approach, originally developed in
\cite{GLPSY} specifically for $\Sigma\Delta$ quantization, relies on
the observation that when the original signal is exactly sparse,
compressed measurements are in fact redundant frame coefficients of
the sparse signal restricted to its support. Since then it has been extended for beta encoding and applied to compressible signals as well \cite{Chou_thesis}. We start with the case of sparse signals.

\subsubsection{Sparse signals}

Let $x\in
\Sigma_k^N$ with $\text{supp}(x)=T$ and $\Phi\in \R^{m\times N}$ be a
compressive sampling matrix. Then, we have
$$y=\Phi x \ \ \implies \ \ y=\Phi_T x_T,$$
where $\Phi_T$ is the submatrix of $\Phi$ consisting of its columns
indexed by $T$ and $x_T$ is the restriction of $x$ to
$T$. Accordingly, \emph{any} quantization technique designed for frames could be
adopted to compressive sampling as follows: \medskip

\noindent {\bf Quantization:} Since the compressive samples are in
fact frame coefficients, apply the noise-shaping quantization algorithm
directly to the compressive samples $y$ to obtain the quantized
samples, say, $q$. Note that the quantization process is blind
to the support of the sparse signal as well as to the sampling operator.  \medskip

\noindent {\bf Reconstruction:} Reconstruct via the following
\emph{two-stage reconstruction algorithm}. To obtain an estimate
$x^{\#}$ of $x$ from $q$:
\begin{enumerate}
\item {\bf Coarse Recovery:} Solve \begin{equation}
    \tilde{x}=\Delta_1^{\epsilon_Q}(\Phi, q)
    \label{eq:l1_SD} 
\end{equation} 
where $\epsilon_Q$ is an upper bound on $\|y-q\|_2$, which depends on
the quantization scheme and is known explicitly. Note that the
decoder $\Delta_1^{\epsilon_Q}$ above can be replaced with any robust
compressive sampling decoder $\Delta$. Clearly, by \eqref{now26}
$\|x-\tilde{x}\|$ will be small if $\epsilon_Q$ is small.
\smallskip

\item {\bf Fine Recovery:} Obtain a support estimate, $\tilde{T}$, of
  $x$ from $\tilde{x}$. A finer approximation for $x$ is then given by
  reconstructing with an appropriate alternative dual
  of the underlying frame $\Phi_{\tilde{T}}$ based on the noise-shaping operator that was employed for quantization.
\end{enumerate}

The success of the two-stage reconstruction algorithm relies on the
accurate recovery of the support of $x$. In turn, this can be
guaranteed by a size condition on the smallest-in-magnitude non-zero
entry of $x$. To see this, note that for all $i \in T$, the robustness guarantee
  \eqref{now26} yields $|\tilde{x}_i - x| \leq C \epsilon_Q$, which,
  together with  the size condition $\min_{i\in T } |x_i| > 2C \epsilon_Q$,
  gives $|\tilde{x}_i|> C \epsilon_Q$. Moreover, by \eqref{now26} we
  have $|\tilde{x}_i| \leq C \epsilon_Q$ for all $i \in
  T^c$. Consequently, the largest-in-magnitude $k$ coefficients of
  $\tilde{x}$ are supported on $T$.  Thus, we have the following proposition.

\begin{proposition}\label{prop:sup_rec}
  Suppose that $x \in \Sigma_k^N$ with $supp(x)=T$, and let $\Phi \in
  \R^{m\times N}$ be a compressive sampling matrix so that
  \eqref{now26} holds for $\Delta = \Delta_1^{\epsilon_Q}$ with robustness constant $C$. Let $\tilde{x}$ be as in \eqref{eq:l1_SD} where $\|\Phi x - q\|_2 \leq \epsilon_Q$. If $\min_{i\in T} |x_i| > 2 C
  \epsilon_Q$, then the $k$ largest-in-magnitude coefficients of
  $\tilde{x}$ are supported on $T$.
\end{proposition}

By this observation, the coarse recovery stage not only yields an
estimate $\tilde{x}$ that satisfies $\|x-\tilde{x}\|_2\leq C
\epsilon_Q$, but it also gives an accurate estimate of the support of
$x$ (via the support of the $k$-largest coefficients of
$\tilde{x}$). It remains to show that reconstruction techniques
associated with noise shaping quantization for frames can be used in
the fine recovery stage to produce an estimate $x^{\#}$ that is more
accurate than $\tilde{x}$ of the coarse stage.

 When $q$ results from a noise-shaping quantization scheme,
 accurate recovery based on alternative duals can be guaranteed via
 \eqref{error_bound_PsiHu}. In particular, suppose that $H$ is
 the noise transfer operator of the quantizer. Conditioned on recovering
 $T$, let $\Psi_{H^{-1}}$ be the left inverse of $\Phi_T$ as defined
 in \eqref{opt-Psi-H} and set $x^{\#}:=
 \Psi_{H^{-1}} q$.  We then have, as before, 
\begin{equation}\label{a-A-H-wtf}
\|x-x^{\#}\|_2 \le \frac{\sqrt{m}}{\sigmin(H^{-1}\Phi_T)}\|u \|_\infty
\end{equation}
where $u$ is as in \eqref{y-q-H-u} .

Predominantly, compressed sensing matrices $\Phi$ (hence their
submatrices $\Phi_T$) are random matrices. Thus, to uniformly control
the reconstruction error via \eqref{a-A-H-wtf} one needs lower bounds
on the smallest singular values of the random matrices $H^{-1}\Phi_T$
for all $T\subset [N]:=\{1,\dots,N\}$, $|T|=k$, as well as a uniform
upper bound on $\|u\|_\infty$.

We concentrate again on random matrices $\Phi$ with independent and identically
distributed Gaussian or sub-Gaussian entries. In these cases, for each fixed support $T$, $\Phi_T$ is a random frame of the type considered in Section~\ref{random-frames} and a probabilistic lower bound
on $\sigma_{\min}(H^{-1}\Phi_T)$ follows from Theorem~\ref{Sobolev_dual_bound} (for Gaussian entries) and Theorem~\ref{thm:main_1} (for sub-Gaussian entries).

A uniform lower bound on
$\sigma_{\min}(H^{-1}\Phi_T)$  over all support
sets $T$ of size $k$ can now be deduced via a union bound over
the ${N}\choose{k}$ support sets.
 Note that to obtain a uniform bound over this rather large set of supports, one requires a relatively small bound for the probability of failure on each potential support, and, consequently, a larger embedding dimension $m$ as compared to the case of a single frame. An alternative approach based on the restricted isometry constant, essentially yielding the same result, can be found in \cite{FK14}.

The approaches just outlined are general and can be applied in the
case of any noise shaping quantizer that allows exact recovery of the
support of sparse vectors via Proposition~\ref{prop:sup_rec}.  In the
following, however, we focus on the special case of $r$th-order $\sd$
quantization, where $H=D^{-r}$ and we obtain the following theorem.

\begin{theorem}[\cite{GLPSY,subGaussian}]\label{thm:polynomial}
Let $r\in \Z^+$, fix $a\in\N$, $\gamma<1$, and $c, C>0$. Then there exist constants $C_{1}, C_{2},
C_{3}, C_{4}$ depending only on these parameters such that the following holds.

Fix $0<\alpha<1$. Let $\Phi$ be an $m \times N$ matrix with
independent sub-Gaussian entries that have zero mean, unit variance, and
parameter $c$, let $\Delta$ be a robust compressive sampling decoder and $k\in\N$
is such that 
  $$\lambda:=\frac{m}{k}\geq \Big(C_{1}  \log(eN/k)\Big)^{\frac{1}{1-\alpha}}.$$

  Suppose that $q$ is obtained by quantizing $\Phi z$, $z\in \R^N$,
  via the $r$th order greedy $\sd$ scheme with the alphabet $\sA_{L,\delta}$, and with $L\geq \lceil\frac{K\lambda^{-1/2}}{\delta}\rceil
  +2^r+1$. Denote by $q$ the quantization output resulting from $\Phi
  z$ where $z\in\R^N$. Then with probability exceeding $1-4e^{-C_{2}
    m^{1-\alpha}k^{\alpha}}$ for all $x\in \Sigma_k^{N}$ having
  $\min\limits_{j\in \rm{supp}{(x)}}|x_j| > C_{3} \delta$:
 \begin{enumerate}
 \item[(i)] The support of $x$, $T$, coincides with the support of the best $k$-term approximation of $\Delta(\frac{1}{\sqrt{m}}\Phi,\frac{1}{\sqrt{m}}q)$. 
 \item[(ii)] Denoting by $\Phi_T$ and $F$ the sub-matrix of $\Phi$ corresponding to the support of $z$ and its $r$th order Sobolev dual respectively, and by $x_T\in\R^k$ the restriction of $x$ to its support, we have
  $$\|x_T-Fq\|_2 \leq C_{4} \lambda^{-\alpha (r-1/2)}\delta.$$ 
 \end{enumerate} 
\end{theorem}

We remark that in Theorem \ref{thm:polynomial}, the requirement that  $L\geq \lceil\frac{K\lambda^{-1/2}}{\delta}\rceil +2^r+1$ ensures stability of the $\sd$ scheme while $\min\limits_{j\in \rm{supp}{(x)}}|x_j| > C_{3}  \delta$ implies accurate support recovery.

\subsubsection{Compressible signals}

The two-stage reconstruction algorithm for sparse signals presented above applies equally well to noise-shaping quantization based on beta encoding as discussed in Section \ref{V-duals}. However, it turns out that for beta encoding there is a more powerful reconstruction algorithm which works for compressible signals as well. 

Let $\Phi$ now be an $m \times N$ compressive sampling matrix, and let $H$ be the $m \times m$ noise transfer operator and $V$ be the $p \times m$ condensation operator as in \eqref{H-V-beta}, where again, for simplicity, we have assumed that $m/p$ is an integer. Note that the associated noise-shaping quantization relation
$$ \Phi x - q = H u$$
implies 
$$ V\Phi x -  Vq =   VH u,$$
hence we may consider $ V\Phi$ as a new condensed measurement matrix and $Vq = V\Phi x +  VHu$ as the corresponding perturbed measurement. As before, 
$$\|VHu\|_2 \leq \|VH\|_{\infty \to 2} \|u\|_\infty 
\leq  \sqrt{p} \beta^{-m/p} \|u\|_\infty ,$$
so that if the greedy quantization rule is stable (i.e., $\|u\|_\infty \leq \delta$), then 
we can set  $\epsilon:= \sqrt{p} \beta^{-m/p} \delta$ and consider the decoder $$(q \mapsto \Delta_1^\epsilon(V\Phi,Vq)).$$
As it follows from the discussion of \eqref{robrec}, if  for some $\alpha > 0$, $\gamma_{2k}:=\gamma_{2k}(\alpha V\Phi)$ is sufficiently small (say less than $1/3$), then we have the estimate 
\begin{equation}\label{compressible_beta_err_bound}
 \|x - \Delta_1^\epsilon(V\Phi,Vq)\|_2 \leq C \alpha \epsilon + D  \frac{\sigma_k(x)_1}{\sqrt{k}},
\end{equation}
where $C$ and $D$ are now absolute constants.

For the random (Gaussian) case, the following result is implied by our discussion above and other tools presented earlier in this paper (for a more detailed derivation of a similar result, see \cite{Chou_thesis}):

\begin{theorem} Let $\Phi$ be an $m \times N$ random matrix whose entries are i.i.d. standard Gaussian variables. Let $x \in \mathbb{R}^N$, $\|x\|_2 \leq 1$, and let $q$ be the result of quantizing the measurements $\Phi x$ with the noise transfer operator $H$ from \eqref{H-V-beta} and the alphabet $\sA_{L,\delta}$ where $\beta + 2\sqrt{N}/\delta \leq L$.
Assume $m \geq p \geq k$ are such that $\lambda':=m/p$ is an integer and  
$$\lambda:= \frac{m}{k} \geq C_1 \lambda' \log N/k$$
for some numerical constant $C_1$.
Let $V$ be the $p \times m$ condensation matrix as in \eqref{H-V-beta} and $\epsilon :=\sqrt{p} \beta^{-m/p} \delta$. Then with probability exceeding
$1 - e^{-p/C'_1}$
for another numerical constant $C'_1$, we have
\[ \|x - \Delta_1^\epsilon(V\Phi, Vq)\|_2 \leq 
C L  \delta\sqrt{p/m}\, \beta^{-m/p}  + D \, \frac{\sigma_k(x)_1}{\sqrt{k}}. 
\]
\end{theorem}

We note that the optimal choice of the auxiliary parameters $p$ and $k$ in the above theorem depends on the success probability as well as further information on the amount of compressibility of $x$. A rule of thumb would be to balance the two error terms above corresponding to quantization error and approximation error. Similarly, the choice of $\beta$, $L$, and $\delta$ can be optimized. For example, if $L\geq 2$ is given and fixed, but $\delta$ is variable, then one would minimize the error bound (over $p$, $k$, $\beta$ and $\delta$) within a given probabilistic guarantee objective and a priori knowledge on $x$.

Finally, we end with the following remark: a recent work \cite{SWY15} shows that it is in fact possible to obtain an approximation from $\sd$ quantized compressive samples that is robust to additive noise and is stable for compressible signals. This approximation is obtained via a \emph{one-stage reconstruction method} based on solving a simple convex optimization problem. Furthermore,  by encoding the quantized measurements via a Johnson- Lindenstrauss dimensionality reducing embedding as in \cite{IS13}, one obtains near-optimal rate-distortion guarantees in the case of sparse signals. For details, see \cite{SWY15}.

\section*{Acknowledgements}
  FK and RS acknowledge support by the German Science Foundation (DFG) in
  the context of the Emmy-Noether Junior Research Group KR 4512/1-1
  ``RaSenQuaSI''. {\"O}Y was funded in part by a Natural Sciences and
  Engineering Research Council of Canada (NSERC) Discovery Grant
  (22R82411), an NSERC Accelerator Award (22R68054) and an NSERC
  Collaborative Research and Development Grant DNOISE II (22R07504).

\bibliographystyle{plain}
\bibliography{CGKSY_Sampta}

\end{document}